\documentclass{article}

\newcommand{\ba}{\begin{eqnarray}}
\newcommand{\ea}{\end{eqnarray}}
\begin{document}             

\title{Spectroscopy of pentaquark states}  
\author{R. Bijker\\
Instituto de Ciencias Nucleares, 
Universidad Nacional Aut\'onoma de M\'exico,\\
A.P. 70-543, 04510 M\'exico, D.F., M\'exico
\and
M.M. Giannini and E. Santopinto \\
Dipartimento di Fisica dell'Universit\`a di Genova, 
I.N.F.N., Sezione di Genova, \\
via Dodecaneso 33, 16164 Genova, Italy}
\date{April 8, 2004}
\maketitle                   

\begin{abstract}
We construct a complete classification of $qqqq\bar{q}$ pentaquark states 
in terms of the spin-flavour $SU(6)$ representations. We find that only some
definite $SU(3)$ representations are allowed, that is singlets, octects,
decuplets, anti-decuplets, 27-plets  and 35-plets. The latter three  contain
exotic states, which cannot be constructed from  three quarks only. 
This complete classification is general and model independent and is useful 
both for model builders and experimentalists. The mass spectrum is obtained 
from a G\"ursey-Radicati type mass formula, whose coefficients have 
been determined previously by a study of $qqq$ baryons. The ground state 
pentaquark, which is identified with the recently observed $\Theta^+(1540)$ 
state, is predicted to be an isosinglet anti-decuplet state. Its parity 
depends on the interplay between the spin-flavour and orbital contributions 
to the mass operator. 
\end{abstract}

\section{Introduction}      

Recently, a baryon with positive strangeness $S=+1$ has been identified 
by several experimental groups 
\cite{nakano,itep,saphir,clas,itep2,hermes,svd}. 
A second exotic baryon with charge $Q=-2$ has also been observed \cite{cern}.
These states are exotic in the sense that they cannot be built up from 
three quarks only as is the case for standard baryons. A state with $S=+1$ 
or $Q=-2$ requires at least a pentaquark configuration of the 
type $qqqq\bar{q}$. 

The possibility and the interest for $S=+1$ baryons (or $Z$ baryons) has
been recorded for many years by the PDG up to 1986, but subsequently it was 
dropped because of lack of clear evidence for their existence. 
However, theoretical interest in exotic baryons has continued both for 
heavy (see \cite{lipkin}) and light quarks (see 
\cite{hogaasen,manohar,chemtob,prasza}).

The experimental interest in pentaquarks was triggered by the work of Diakonov
et al. \cite{diakonov}, who predicted an exotic $S=+1$ baryon with a
definite mass and a small width, thus providing an invaluable guide for 
experimentalists. Such a state, the now famous $\Theta^{+}$, is the 
isoscalar member of a flavour anti-decuplet, whose relative 
energies are evaluated by means of a $SU_{\rm f}(3)$ violating
interaction  based on the Skyrme model. The energy scale is fixed
identifying the nucleon-like state with $S=0$ of the anti-decuplet with 
the well-known $N(1710)$ resonance. In this way the obtained value of the 
spin and parity of the $\Theta^{+}$ is $\frac{1}{2}^{+}$. 
However, from the experimental point 
of view, the known properties of $\Theta^{+}$ are: the mass (in remarkably 
coincidence with the prediction of \cite{diakonov}), the width (smaller 
than the one of other $N^*$ resonances of comparable mass, in qualitative 
agreement with the prediction \cite{diakonov}), the strangeness ($S=+1$) 
and the charge ($Q=+1$). Moreover, it seems to be an isosinglet \cite{saphir}. 
In this way it can be safely identified with the isoscalar state of the 
anti-decuplet. On the contrary, the spin and the parity still have to be 
determined.

The discovery of the pentaquark has produced a strongly increased theoretical
interest, giving rise to a long series of papers which address various 
aspects of pentaquarks. Besides the Skyrme model 
\cite{prasza,diakonov,borisyuk,skyrmion,jennings}, 
there are many studies based on the Constituent Quark Model (CQM) 
\cite{stancu,helminen,capstick,hosaka,carlson,glozman,williams,anti10},   
the diquark-diquark-$\bar{q}$ approach \cite{diquark}, 
QCD sum rules \cite{sumrule}, large $N_c$ QCD \cite{largenc}, 
lattice QCD \cite{lattice}, and many others \cite{other}.  
In many cases the models assume 
or predict a definite parity for the $\Theta^{+}$, which in most cases is 
positive \cite{diakonov,borisyuk,stancu,hosaka,glozman,diquark}. 
However, recent work on QCD-sum rules \cite{sumrule} 
and lattice QCD \cite{lattice} implies a negative parity.  

In this article, we study the classification scheme of pentaquark 
states from symmetry principles, leading to a complete basis for the 
$qqqq\bar{q}$ states in terms of the spin-flavour $SU_{\rm sf}(6)$ 
multiplets. Next we calculate the energies of exotic 
pentaquark states using a G\"ursey-Radicati type mass formula, discuss 
some general features of the pentaquark spectrum, and finally address 
the properties of the ground state pentaquark state.  

\section{The classification of pentaquark states}

As for all multiquark systems, the pentaquark wave function contains  
contributions connected to the spatial degrees of freedom  
and the internal degrees of freedom of colour, flavour and spin. 
In order to classify the corresponding states, we shall make use as 
much as possible of symmetry principles without, for the moment, 
introducing any explicit dynamical model. In the construction of the 
classification scheme we are guided by two conditions: the pentaquark 
wave function should be a colour singlet as all physical states, and 
should be antisymmetric under any permutation of the   four quarks. 

We shall make use of the Young tableau technique to 
construct the allowed $SU_{\rm sf}(6)$ representations for the pentaquark 
$q^4\bar{q}$ system, denoting with a box the fundamental representation 
of $SU(n)$, with $n=2$, 3, 6 for the spin, flavour (or colour), and 
spin-flavour degrees of freedom, respectively.
The quark transforms as the fundamental representation $[1]$ under $SU(n)$, 
whereas the antiquark transforms as the conjugate representation 
$[1^{n-1}]$ under $SU(n)$. The spin-flavour classification for the quark 
and antiquark are given by 
\ba
\begin{array}{cccccc}
& SU_{\rm sf}(6) & \supset & SU_{\rm f}(3) & \otimes & SU_{\rm s}(2) \\
& & & & & \\ 
\mbox{quark} & $[1]$ & \supset & $[1]$ & \otimes & $[1]$ \\
& & & & & \\ 
& \setlength{\unitlength}{1.0pt}
\begin{picture}(10,10)(0,0)
\thinlines
\put ( 0, 0) {\line (1,0){10}}
\put ( 0,10) {\line (1,0){10}}
\put ( 0, 0) {\line (0,1){10}}
\put (10, 0) {\line (0,1){10}}
\end{picture} & \supset & 
\setlength{\unitlength}{1.0pt}
\begin{picture}(10,10)(0,0)
\thinlines
\put ( 0, 0) {\line (1,0){10}}
\put ( 0,10) {\line (1,0){10}}
\put ( 0, 0) {\line (0,1){10}}
\put (10, 0) {\line (0,1){10}}
\end{picture} & \otimes &
\setlength{\unitlength}{1.0pt}
\begin{picture}(10,10)(0,0)
\thinlines
\put ( 0, 0) {\line (1,0){10}}
\put ( 0,10) {\line (1,0){10}}
\put ( 0, 0) {\line (0,1){10}}
\put (10, 0) {\line (0,1){10}}
\end{picture} \\
& & & & & \\
\mbox{antiquark} & $[11111]$ & \supset & $[11]$ & \otimes & $[1]$ \\
& & & & & \\
& \setlength{\unitlength}{1.0pt}
\begin{picture}(10,10)(0,0)
\thinlines
\put ( 0, 10) {\line (1,0){10}}
\put ( 0,  0) {\line (1,0){10}}
\put ( 0,-10) {\line (1,0){10}}
\put ( 0,-20) {\line (1,0){10}}
\put ( 0,-30) {\line (1,0){10}}
\put ( 0,-40) {\line (1,0){10}}
\put ( 0,-40) {\line (0,1){50}}
\put (10,-40) {\line (0,1){50}}
\end{picture} & \supset & 
\setlength{\unitlength}{1.0pt}
\begin{picture}(10,10)(0,0)
\thinlines
\put ( 0, 10) {\line (1,0){10}}
\put ( 0,  0) {\line (1,0){10}}
\put ( 0,-10) {\line (1,0){10}}
\put ( 0,-10) {\line (0,1){20}}
\put (10,-10) {\line (0,1){20}}
\end{picture} & \otimes &
\setlength{\unitlength}{1.0pt}
\begin{picture}(10,10)(0,0)
\thinlines
\put ( 0, 0) {\line (1,0){10}}
\put ( 0,10) {\line (1,0){10}}
\put ( 0, 0) {\line (0,1){10}}
\put (10, 0) {\line (0,1){10}}
\end{picture} \\
& & & & & \\
\end{array}
\label{qqbar}
\ea

\

\

\

\noindent
on the right hand we have used inner products of single quark states. The
spin-flavour states of multiquark systems can be obtained by taking the outer 
product of the representations of the quarks and/or antiquarks. 

\subsection{The $q^3$ system}

In order to establish the notation, we start by considering the well-known 
example of $qqq$ baryons. The allowed $SU_{\rm sf}(6)$ states are obtained 
by means of the product
\ba
\setlength{\unitlength}{1.0pt}
\begin{picture}(10,10)(0,0)
\thinlines
\put ( 0, 0) {\line (1,0){10}}
\put ( 0,10) {\line (1,0){10}}
\put ( 0, 0) {\line (0,1){10}}
\put (10, 0) {\line (0,1){10}}
\end{picture} \;\;\otimes\;\; 
\setlength{\unitlength}{1.0pt}
\begin{picture}(10,10)(0,0)
\thinlines
\put ( 0, 0) {\line (1,0){10}}
\put ( 0,10) {\line (1,0){10}}
\put ( 0, 0) {\line (0,1){10}}
\put (10, 0) {\line (0,1){10}}
\end{picture} \;\;\otimes\;\; 
\setlength{\unitlength}{1.0pt}
\begin{picture}(10,10)(0,0)
\thinlines
\put ( 0, 0) {\line (1,0){10}}
\put ( 0,10) {\line (1,0){10}}
\put ( 0, 0) {\line (0,1){10}}
\put (10, 0) {\line (0,1){10}}
\end{picture} &=& 
\setlength{\unitlength}{1.0pt}
\begin{picture}(30,10)(0,0)
\thinlines
\put ( 0, 0) {\line (1,0){30}}
\put ( 0,10) {\line (1,0){30}}
\put ( 0, 0) {\line (0,1){10}}
\put (10, 0) {\line (0,1){10}}
\put (20, 0) {\line (0,1){10}}
\put (30, 0) {\line (0,1){10}}
\end{picture} \;\;\oplus\;\; 2 \;\;  
\setlength{\unitlength}{1.0pt}
\begin{picture}(20,20)(0,5)
\thinlines
\put ( 0, 0) {\line (1,0){10}}
\put ( 0,10) {\line (1,0){20}}
\put ( 0,20) {\line (1,0){20}}
\put ( 0, 0) {\line (0,1){20}}
\put (10, 0) {\line (0,1){20}}
\put (20,10) {\line (0,1){10}}
\end{picture} \;\;\oplus\;\; 
\setlength{\unitlength}{1.0pt}
\begin{picture}(10,30)(0,10)
\thinlines
\put ( 0, 0) {\line (1,0){10}}
\put ( 0,10) {\line (1,0){10}}
\put ( 0,20) {\line (1,0){10}}
\put ( 0,30) {\line (1,0){10}}
\put ( 0, 0) {\line (0,1){30}}
\put (10, 0) {\line (0,1){30}}
\end{picture} 
\ea

\

\noindent
In the following, we adopt for the representations the notation 
$[f]_d=[f_1,\ldots,f_n]_d$, where $f_i$ denotes the number of boxes in 
the $i$-th row of the Young tableau, and $d$ is the dimension of the 
representation. In this way, the above product is written as 
\ba
[1]_6 \;\otimes\; [1]_6 \;\otimes\; [1]_6 &=& 
[3]_{56} \;\oplus\; 2 \, [21]_{70} \;\oplus\; [111]_{20} ~. 
\label{qqqsu6}
\ea
In Table~\ref{qqq} we summarize the results for the allowed spin-flavour, 
flavour (colour) and spin states of $q^3$ baryons. The spin states are 
given by the representations $[f_1f_2]=[30]$ and $[21]$ or, equivalently, 
by their spin $s=(f_1-f_2)/2=3/2$ and $1/2$, respectively. On the left-hand 
side we show the labels of the point group $D_3$ which is isomorphic to the   
permutation group of three identical objects $S_3$. 
A complete classification of three quark states involves the analysis of 
the flavour and spin content of each spin-flavour representation, i.e. the 
decomposition of representations of $SU_{\rm sf}(6)$ into those of 
$SU_{\rm f}(3) \otimes SU_{\rm s}(2)$ (see also Table~\ref{sfqqq})  
\ba
\, [3]_{56} &÷=÷& ([21]_8 \;\otimes\; [21]_2) 
\;\oplus\; ([3]_{10} \;\otimes\; [3]_4) ~,  
\nonumber\\
\, [21]_{70} &÷=÷& ([21]_8 \;\otimes\; [21]_2) 
\;\oplus\; ([21]_8 \;\otimes\; [3]_4) 
\;\oplus\; ([3]_{10} \;\otimes\; [21]_2) 
\;\oplus\; ([111]_1 \;\otimes\; [21]_2) ~,
\nonumber\\
\, [111]_{20} &÷=÷& ([21]_8 \;\otimes\; [21]_2) 
\;\oplus\; ([111]_1 \;\otimes\; [3]_4) ~,
\label{qqq1}
\ea
or in the usual notation
\ba
\, [56] &÷=÷& ^{2}8 \;\oplus\; ^{4}10 ~,
\nonumber\\
\, [70] &÷=÷& ^{2}8 \;\oplus\; ^{4}8 \;\oplus\; ^{2}10  
\;\oplus\; ^{2}1 ~,
\nonumber\\
\, [20] &÷=÷& ^{2}8 \;\oplus\; ^{4}1 ~.
\label{qqq2}
\ea

\subsection{The $q^4$ system}

To study the structure of pentaquark $q^4\bar{q}$ states, it is convenient 
to first construct the $qqqq$ states which should satisfy Pauli statistics, 
and then to add the $\bar{q}$ antiquark. \\
The allowed $SU_{\rm sf}(6)$ 
spin-flavour states of the $q^4$ system follow from the product of the 
$q^3$ configurations of Eq.~(\ref{qqqsu6}) and a single quark 
\ba
\, [3]_{56} \;\otimes\; [1]_6 &=& [4]_{126} \;\oplus\; [31]_{210} ~,
\nonumber\\
\, [21]_{70} \;\otimes\; [1]_6 &=& [31]_{210} \;\oplus\; [22]_{105} 
\;\oplus\; [211]_{105} ~,
\nonumber\\
\, [111]_{20} \;\otimes\; [1]_6 &=& [211]_{105} \;\oplus\; [1111]_{15} ~. 
\ea
As a result we obtain for the $q^4$ spin-flavour states 
\ba
[1]_6 \;\otimes\; [1]_6 \;\otimes\; [1]_6 \;\otimes\; [1]_6 &=& 
[4]_{126} \;\oplus\; 3 \, [31]_{210} \;\oplus\; 2 \, [22]_{105} 
\;\oplus\; 3 \, [211]_{105} \;\oplus\; [1111]_{15} ~.
\label{qqqqsu6}
\ea
In Table~\ref{qqqq} we summarize the results for the allowed spin-flavour, 
flavour (colour) and spin states of a system of four identical quarks. The 
permutation symmetry is characterized by the $S_4$ Young tableaux $[4]$, 
$[31]$, $[22]$, $[211]$ and $[1111]$ or, equivalently, by the irreducible 
representations of the tetrahedral group ${\cal T}_d$ (which is isomorphic 
to $S_4$) as $A_1$, $F_2$, $E$, $F_1$ and $A_2$, respectively. The flavour 
and spin content of the various $q^4$ configurations of Eq.~(\ref{qqqqsu6}) 
is presented in Table~\ref{sfqqqq}. The ${\cal T}_d$ labels denote the 
permutation symmetry of the four-quark system, and the $D_3$ labels that 
of the three-quark subsystem.  

\subsection{The $q^4 \bar{q}$ system}

The pentaquark configurations are now obtained by considering the product 
of the $q^4$ states of Eq.~(\ref{qqqqsu6}) and the antiquark state of 
Eq.~(\ref{qqbar}). The allowed $SU_{\rm sf}(6)$ states are 
\ba
\, [4]_{126} \;\otimes\; [11111]_6 &=& [51111]_{700} 
\;\oplus\; [411111]_{56} ~,
\nonumber\\
\, [31]_{210} \;\otimes\; [11111]_6 &=& [42111]_{1134} 
\;\oplus\; [411111]_{56} \;\oplus\; [321111]_{70} ~,
\nonumber\\
\, [22]_{105} \;\otimes\; [11111]_6 &=& [33111]_{560} 
\;\oplus\; [321111]_{70} ~, 
\nonumber\\
\, [211]_{105} \;\otimes\; [11111]_6 &=& [32211]_{540} 
\;\oplus\; [321111]_{70} \;\oplus\; [222111]_{20} ~, 
\nonumber\\
\, [1111]_{15} \;\otimes\; [11111]_6 &=& [22221]_{70} 
\;\oplus\; [222111]_{20} ~, 
\ea
As a result, we obtain for the $q^4\bar{q}$ spin-flavour states 
\ba
[1]_6 \;\otimes\; [1]_6 \;\otimes\; [1]_6 \;\otimes\; [1]_6 
\;\otimes\; [11111]_6 &=& [51111]_{700} \;\oplus\; 4 \, [411111]_{56} 
\;\oplus\; 3 \, [42111]_{1134} 
\nonumber\\
&& \;\oplus\; 8 \, [321111]_{70} \;\oplus\; 2 \, [33111]_{560} 
\;\oplus\; 3 \, [32211]_{540} 
\nonumber\\
&& \;\oplus\; 4 \, [222111]_{20} \;\oplus\; [22221]_{70} ~.
\label{pentasu6}
\ea
In a similar way, one can construct the allowed flavour multiplets as 
\ba
[1]_3 \;\otimes\; [1]_3 \;\otimes\; [1]_3 \;\otimes\; [1]_3 
\;\otimes\; [11]_3 &=& [51]_{35} \;\oplus\; 3 \, [42]_{27} 
\;\oplus\; 2 \, [33]_{\overline{10}} 
\nonumber\\
&& \;\oplus\; 4 \, [411]_{10} \;\oplus\; 8 \, [321]_{8} 
\;\oplus\; 3 \, [222]_{1} ~.
\label{pentasu3}
\ea
The allowed spin states are obtained from 
\ba
[1]_2 \;\otimes\; [1]_2 \;\otimes\; [1]_2 \;\otimes\; [1]_2 
\;\otimes\; [1]_2 &=& [5]_{6} \;\oplus\; 4 \, [41]_{4} 
\;\oplus\; 5 \, [32]_{2} ~,
\label{pentasu2}
\ea
where the configurations $[5]$, $[41]$ and $[32]$ correspond to the spin 
values $s=5/2$, $3/2$ and $1/2$, respectively. 
In Table~\ref{states}, we summarize the results for the allowed spin, 
flavour and spin-flavour states for $q^4 \bar{q}$ pentaquarks. 
The ${\cal T}_d$ labels in the last column denote the permutation 
symmetry of the four-quark subsystem. 

The full decomposition of the spin-flavour states of Eq.~(\ref{pentasu6}) 
into the spin and flavour states of Eqs.~(\ref{pentasu2}) and~(\ref{pentasu3}) 
is presented in Table~\ref{pentaquark}. The results are in agreement 
with the reduction of the colour-spin $SU_{\rm cs}(6)$ algebra of 
\cite{hogaasen}. The spin and flavour content of the $SU_{\rm sf}(6)$ 
representations $[411111]_{56}$, $[321111]_{70}$ and $[222111]_{20}$ 
is the same as that of the representations $[3]_{56}$, $[21]_{70}$ and 
$[111]_{20}$ for the three-quark system in Eqs.~(\ref{qqq1}) 
and~(\ref{qqq2}). This means that the states belonging to 
these representations have the same quantum numbers as the  
$qqq$ system and hence are difficult to distinguish from the commonly
known baryon resonances. Therefore, exotic states, that is pentaquarks
having quantum numbers not obtainable with three-quark configurations,
are to be looked for in the remaining five $SU_{\rm sf}(6)$ 
representations of Eq.~(\ref{pentasu6}): $[51111]_{700}$, 
$[42111]_{1134}$, $[33111]_{560}$, $[32211]_{540}$ and $[22221]_{70}$. 
Their decomposition into spin and flavour states can be found in 
Table~\ref{pentaquark}. In the notation of Eq.~(\ref{qqq2}), 
we can write
\begin{eqnarray}
\, [700] &=& ^{2}8 \;\oplus\; ^{4}8 \;\oplus\; ^{2}10 \;\oplus\; ^{4}10 
\;\oplus\; ^{6}10 
\nonumber\\
&& \;\oplus\; ^{2}\overline{10} \;\oplus\; ^{2}27 
\;\oplus\; ^{4}27 \;\oplus\; ^{4}35 \;\oplus\; ^{6}35 ~,
\nonumber\\
\, [1134] &=& ^{2}1 \;\oplus\; ^{4}1 \;\oplus\; 3( ^{2}8) 
\;\oplus\; 3( ^{4}8) \;\oplus\; ^{6}8 \;\oplus\; 2( ^{2}10) 
\;\oplus\; 2( ^{4}10) \;\oplus\; ^{6}10 
\nonumber\\ 
&& \;\oplus\; ^{2}\overline{10} \;\oplus\; ^{4}\overline{10} 
\;\oplus\; 2( ^{2}27) \;\oplus\; 2( ^{4}27) \;\oplus\; ^{6}27
\;\oplus\; ^{2}35 \;\oplus\; ^{4}35 ~.
\nonumber\\
\, [560] &=& ^{4}1 \;\oplus\; 2( ^{2}8) \;\oplus\; 2( ^{4}8) 
\;\oplus\; ^{6}8 \;\oplus\; ^{2}10 \;\oplus\; ^{4}10 
\nonumber\\
&& \;\oplus\; ^{2}\overline{10} \;\oplus\; ^{4}\overline{10} 
\;\oplus\; ^{6}\overline{10} \;\oplus\; ^{2}27 \;\oplus\; ^{4}27 
\;\oplus\; ^{2}35 ~,
\nonumber\\
\, [540] &=& ^{2}1 \;\oplus\; ^{4}1 \;\oplus\; ^{6}1  \;\oplus\; 3( ^{2}8) 
\;\oplus\; 3( ^{4}8) \;\oplus\; ^{6}8 \;\oplus\; ^{2}10 \;\oplus\; ^{4}10 
\nonumber\\
&& \;\oplus\; ^{2}\overline{10} \;\oplus\; ^{4}\overline{10} 
\;\oplus\; 2 ( ^{2}27) \;\oplus\; ^{4}27 ~, 
\nonumber\\
\, [\overline{70}] &=& ^{2}1 \;\oplus\; ^{2}8 \;\oplus\; ^{4}8 
\;\oplus\; ^{2}\overline{10} ~, 
\label{exo}
\end{eqnarray}
It is difficult to distinguish the pentaquark flavour singlets, 
octets and decuplets from the standard three-quark states. 
The $SU_{\rm f}(3)$ representations $\overline{10}$, $27$ and $35$ 
(see Figs.~\ref{flavour33e}--\ref{flavour51a1}) contain exotic 
states which cannot be obtained from three-quark configurations only.
These states are more easily identified experimentally because of the 
uniqueness of their quantum numbers.  
In Table~\ref{exotic} we present a complete list of exotic pentaquark 
states. For each isospin multiplet we have identified the states whose 
combination of hypercharge $Y$ and charge $Q$ cannot be obtained with 
three-quark configurations. In Figs.~\ref{flavour33e}--\ref{flavour51a1} 
the exotic states are indicated by $\bullet$. 

So far, we have discussed the spin-flavour part of the pentaquark wave 
function with $S_4$ (or ${\cal T}_d$ symmetry). The spin-flavour part 
has to be combined  
with the colour part and the orbital (or radial) part in such a way 
that the total pentaquark wave function is a $[222]_1$ colour-singlet 
state, and that the four quarks obey the Pauli principle, i.e. are 
antisymmetric under any permutation of the four quarks. 
Since the colour part of the pentaquark wave function is a $[222]_1$ singlet 
and that of the antiquark a $[11]_3$ anti-triplet, 
the colour wave function of the four-quark configuration is a $[211]_{3}$ 
triplet with $F_1$ symmetry under ${\cal T}_d$. 
The total $q^4$ wave function is antisymmetric ($A_2$), hence the 
orbital-spin-flavour part is a $[31]$ state with $F_2$ symmetry 
which is obtained from the colour part by interchanging rows and columns 
\ba
\psi_{\rm c}(q^4) \;&\; [211] \;&\; F_1 
\nonumber\\
\psi_{\rm osf}(q^4) \;&\; [31] \;&\; F_2
\ea

Next we discuss the symmetry properties of the orbital part of the 
pentaquark wave function. If the four quarks are in a spatially 
symmetric $S$-wave ground state with $A_1$ symmetry, the only allowed 
$SU_{\rm sf}(6)$ representation is $[31]$ with $F_2$ symmetry. 
According to Table~\ref{pentaquark}, the only pentaquark configuration 
with $F_2$ symmetry that contains exotic states is $[42111]_{1134}$. 
On the other hand, if the four quarks are in a $P$-wave state with $F_2$ 
symmetry, there are several allowed $SU_{\rm sf}(6)$ representations: 
$[4]$, $[31]$, $[22]$ and $[211]$ with $A_1$, $F_2$, $E$ and $F_1$ 
symmetry, respectively. The corresponding pentaquark configurations 
that contain exotic states are $[51111]_{700}$, $[42111]_{1134}$, 
$[33111]_{560}$ and 
$[32211]_{540}$, respectively. In Table~\ref{orbital} we present for each 
symmetry type of the orbital wave function, the corresponding symmetry 
of the spin-flavour wave function, as well as the associated pentaquark 
configurations that contain exotic states. 
The explicit construction of the $S_4$ invariant orbital-spin-flavour 
pentaquark wave functions will be presented in a separate publication 
\cite{BGS}. The methods are analogous to those used for the $S_3$ 
invariant $qqq$ baryon wave functions (see e.g. \cite{FKR,ik,bil,sig}). 

We would like to stress the general validity of these results.  
The classification scheme derived in this section is complete, and is 
based only on the fact that quarks (and antiquarks) have orbital, colour, 
spin and flavour degrees of freedom. The precise ordering of the pentaquark 
states in the mass spectrum depends on the choice of a specific dynamical 
model (Skyrme, CQM, Goldstone Boson Exchange, instanton, hypercentral, 
stringlike, ...). In the case of the Skyrmion model many states are 
suppressed because of a strict correlation between spin and isospin.

\section{The pentaquark spectrum}

In order to study the general structure of the spectrum of exotic 
pentaquarks, we consider a simple schematic model in which the mass 
operator is given by 
\ba
M \;=\; M_0 + M_{\rm orb} + M_{\rm sf} ~. 
\label{mass}
\ea
$M_0$ is a constant. $M_{\rm orb}$ describes the contribution to the 
pentaquark mass due to the space degrees of freedom of the constituent 
quarks. The last term $M_{\rm sf}$ contains the spin-flavour dependence 
and is assummed to have a generalized G\"ursey-Radicati form 
\ba
M_{\rm sf} &=& -A \, C_{2SU_{\rm sf}(6)} 
+ B \, C_{2SU_{\rm f}(3)} + C \, s(s+1)
\nonumber\\
&& + D \, Y + E \, [I(I+1)-\frac{1}{4}Y^2] ~. 
\label{grmass}
\ea
The first two terms represent the quadratic Casimir operators of the 
$SU_{\rm sf}(6)$ spin-flavour and the $SU_{\rm f}(3)$ flavour groups,  
and $s$, $Y$ and $I$ denote the spin, hypercharge and isospin, respectively. 
For the definition of the Casimir operators in Eq.~(\ref{grmass}), we have 
followed the same convention as in \cite{helminen}. 
The eigenvalues of the Casimirs are given by 
\ba
C_{2SU_{\rm sf}(n)} &=& \frac{1}{2} \left[ \sum_{i=1}^n f_i(f_i+n+1-2i) 
- \frac{1}{n} \left(\sum_{i=1}^n f_i\right)^2 \right] ~.
\label{cas}
\ea
In Table~\ref{casimir}, we give the expectation 
values of the Casimir operators $C_{2SU_{\rm sf}(6)}$ and 
$C_{2SU_{\rm f}(3)}$ for the allowed pentaquark configurations. 

The last two terms in Eq.~(\ref{grmass}) correspond to the Gell-Mann-Okubo 
mass formula that describes the splitting within a flavour multiplet 
\cite{GMO}. This formula was extended by G\"ursey and Radicati \cite{GR} 
to include the terms proportional to $B$ and $C$ that depend on the spin 
and the flavour representations, which in turn was generalized further to 
include the spin-flavour term proportional to $A$ as well \cite{bil}. 

In many studies of multiquark configurations, effective spin-flavour 
hyperfine interactions have been used in CQM which schematically 
represents the Goldstone Boson Exchange (GBE) interaction between 
constituent quarks \cite{stancu,helminen,carlson,glozman}. 
An analysis of the strange and non-strange $qqq$ baryon resonances in 
the collective stringlike model \cite{bil} and the hypercentral CQM 
\cite{gsv} also showed evidence for the need of such type of interaction 
terms. 
If one neglects their radial dependence, the matrix elements of these 
interactions depend on the Casimirs of the $SU_{\rm sf}(6)$ spin-flavour, 
the $SU_{\rm f}(3)$ flavour and the $SU_{\rm s}(2)$ spin groups  
\cite{helminen}
\ba
\left< \sum_{i<j}^n (\vec{\lambda}_i \cdot \vec{\lambda}_j) 
(\vec{\sigma}_i \cdot \vec{\sigma}_j) \right> \;=\; 
4 C_{2SU_{\rm sf}(6)} - 2 C_{2SU_{\rm f}(3)} 
- \frac{4}{3} s(s+1) - 8 n ~.
\label{hyperfine}
\ea
\noindent where $n$ is the number of quarks. 

The energy splittings within a given multiplet induced by Eq.~(\ref{hyperfine}) 
have the same structure as the G\"ursey-Radicati formula of Eq.~(\ref{grmass}), 
with the exception of the Gell-Mann-Okubo term. The constant with the number of
quarks cancels out  when evaluating energy differences. The dependence on the 
different quark numbers is taken into account by the fact that the eigenvalues 
of the Casimirs for the $qqq$ or $qqqq\bar{q}$ states can be very different. 
The interaction of Eq.~(\ref{hyperfine}) is not the most general one. For instance, 
the presence of an explicit spin-spin interaction would 
modify the $-4/3$ coefficient. 

In Eqs.~(\ref{mass}) and~(\ref{grmass}) we have made a very strong approximation: 
we have neglected the spatial dependence of the $SU_{sf}(6)$ breaking part. 
As a consequence, there is no $SU_{\rm sf}(6)$ mixing. The kind of problems that 
can arise neglecting the spatial dependence in the $SU_{sf}(6)$ breaking interaction 
is discussed by Jennings and Maltman \cite{jennings} for two of the models in the 
literature, the Goldstone boson model and the bag model. 

The average energy of $SU_{\rm sf}(6)$ multiplets depends on the orbital part 
$M_{\rm orb}$ and on the term linear in the $SU_{\rm sf}(6)$ Casimir, 
while the terms proportional to $B$, $C$, $D$ and $E$ give the splittings inside 
the multiplet. At the moment, there is experimental evidence for two 
pentaquark states. This is not sufficient to determine all parameters in 
the mass formula, and then to predict the masses of other pentaquarks.  
For this reason we use the values of the parameters determined from the three-quark 
spectrum, assuming that the coefficients in the GR are the same for different quark 
systems. Clearly, new experimental data on the pentaquark states will allow to 
determine how different can be the parameters relevant
for the pentaquark spectrum with respect to the $qqq$ ones.
 
In the case of the $qqq$ system, the coefficients $B$, $C$, $D$ and $E$ can be 
obtained from the mass
differences of selected pairs of baryon resonances \cite{gsv}
\ba
M_{\Delta(1232)} - M_{N(938)} &=& 3(B+C+E) ~, 
\nonumber\\
M_{N(1650)} - M_{N(1535)} &=& 3C ~, 
\nonumber\\
4M_{N(938)} - M_{\Sigma(1193)} - 3M_{\Lambda(1116)} &=& 4D ~,
\nonumber\\
M_{\Sigma(1193)} - M_{\Lambda(1116)} &=& 2E ~, 
\ea
leading to the numerical values
\ba
\begin{array}{lcrl}
B &=&  21.2 & \mbox{MeV} ~, \\
C &=&  38.3 & \mbox{MeV} ~, \\
D &=&-197.3 & \mbox{MeV} ~, \\
E &=&  38.5 & \mbox{MeV} ~.
\end{array}
\label{grpar}
\ea
The coefficients we have so obtained can be used for a preliminary evaluation of the 
splittings within any $SU_{\rm sf}(6)$ multiplet, assuming that they do not depend on 
the quark system, just as is the case for the hyperfine interaction of Eq.~(\ref{hyperfine}). 
The eigenvalues of the Casimirs for the $qqq$ or $qqqq\bar{q}$ systems
are different (see Table \ref{casimir}) and in this way the
presence of a different quark structure is taken into account. 

We use the G\"ursey-Radicati formula  for the calculation of the energy splittings 
of the exotic pentaquark states, using the constant $M_0$ in order to normalize the 
energy scale to the observed mass of the $\Theta^{+}$.
The results are shown in Table \ref{pentamass1}, where neither the $M_{orb}$ nor the 
$A~ C_{2SU_{\rm sf}(6)}$ terms have been introduced. 
Table~\ref{pentamass1} shows that for all spin-flavour 
configurations the lowest pentaquark state is characterized by 
$^{2}\overline{10}$, i.e. a flavour anti-decuplet $[33]$ state with 
spin $s=1/2$ and isospin $I=0$, in agreement with the available 
experimental data which indicate that the $\Theta^+(1540)$ is 
an isosinglet \cite{saphir}. 
For all spin-flavour configurations, there are other lowlying excited 
pentaquark states belonging to the 27-plet at 1660 MeV and 1775 MeV. 
The anti-decuplet state with strangeness $S=-2$ ($Y=-1$) and isospin 
$I=3/2$ is calculated at an energy of 2305 MeV, to be compared with 
the recently observed resonance at 1862 MeV \cite{cern} which was suggested 
as a candidate for the $\Xi_{3/2}$ exotic with charge $Q=-2$.

Another 
important consequence of the use of a `diagonal' form of the interactions 
in Eq.~(\ref{grmass}) is that the structure of the wave functions does 
not depend on the values of the coefficients. A change in the coefficients  
causes a shift in the energies, but does not modify the wave functions.

The degeneracy of the multiplets in Table~\ref{pentamass1} can be eliminated if one
considers the contributions from the Casimir of $SU(6)$ and from the space term
$M_{\rm orb}$. For the consistent treatment of the latter one needs a specific model,
but this is beyond the scope of this work. Nevertheless, we shall present some
general arguments in the next section which are relevant for the spin and parity 
of the ground state pentaquark. Here we concentrate ourselves on the effects of the term
linear in $A$ in Eq.~(\ref{grmass}) on the energy splittings of pentaquark states. The
value of the coefficient $A$ can be determined, analogously to what has been done in
connection with Eq.~(\ref{grpar}), from the energy difference between the lowest $S_{11}$
resonance and the Roper
\ba
  M_{N(1535)} - M_{N(1440)} \;=\; 3A + \Delta M_{\rm orb} ~. 
\ea
$\Delta M_{\rm orb}$ is the orbital contribution to the mass difference, and can be
taken from the $SU(6)$ invariant energies provided by the HCQM \cite{bil,gsv}, which 
leads to a value of $A=55.1$ MeV. The positive sign of $A$ is in agreement with the 
sign used in previous studies of baryons as $qqq$ configurations \cite{bil}.

In Table~\ref{pentamass2} we present the spin-flavour contribution to the 
energies of all exotic pentaquark states for the four allowed 
$SU_{\rm sf}(6)$ spin-flavour multiplets. The effect of the spin-flavour 
term shifts the different $SU_{\rm sf}(6)$ 
multiplets with respect to one another, without changing 
their internal structure. The lowest pentaquark state has the labels $^{2}\overline{10}$,  
i.e. is an anti-decuplet state with spin $s=1/2$ and 
isospin $I=0$, belonging to the $[51111]_{700}$ multiplet. 
The parity of this state is positive. 

In the next section, we discuss the effect of orbital excitation energies 
on the angular momentum and parity of the ground state pentaquark. It is 
important to note that, irrespective of the orbital contribution to the mass, 
the ground state pentaquark is an anti-decuplet flavour state with spin 
$s=1/2$ and isospin $I=0$.

\subsection{Spin and parity of the ground state pentaquark}

What are the consequences of these calculations for the spin and parity 
of the $\Theta^+(1540)$? This depends in part on the assignment of 
quantum numbers, and in part on the choice of a particular model to 
describe the orbital motion. In the following we identify the  
$\Theta^+(1540)$ resonance with the ground state exotic pentaquark 
configuration.

The treatment of the orbital part is very much dependent on the choice 
of a specific dynamical model (harmonic oscillator, Skyrme, soliton, 
stringlike, hypercentral, ...). We consider a simple model 
in which the orbital motion of the pentaquark is limited to excitations 
up to $N=1$ quantum. The model space consists of five states: an 
$S$-wave state with $L^p=0^+$ and $A_1$ symmetry for the four quarks, 
and four excited $P$-wave states with $L^p=1^-$, three of which 
correspond to excitations in the relative coordinates of the four-quark 
subsystem and the fourth to an excitation in the relative coordinate 
between the four-quark subsystem and the antiquark. As a consequence of 
the $S_4$ permutation symmetry of the four quarks, the first three 
excitations form a degenerate triplet with three-fold $F_2$ symmetry, 
and the fourth has $A_1$ symmetry. 
In summary, the states in this simple model for the orbital motion are 
characterized by angular momentum $L$, parity $p$ and ${\cal T}_d$ 
symmetry $t$: $L^p_t=0^+_{A_1}$, $1^-_{F_2}$ and $1^-_{A_1}$. 
The total angular momentum of the pentaquark state is given by 
$\vec{J}=\vec{L}+\vec{s}$, whereas the parity is opposite 
to that of the orbital excitation due to the negative intrinsic 
parity of the $q^4 \bar{q}$ configuration. 
According to Table~\ref{orbital}, the exotic spin-flavour states associated 
with the orbital states $L^p_t=0^+_{A_1}$ and $1^-_{A_1}$ belong to 
the $[42111]_{1134}$ representation, whereas the state $L^p_t=1^-_{F_2}$ 
gives rise to exotic pentaquark states belonging to the $[51111]_{700}$, 
$[42111]_{1134}$, $[33111]_{560}$ and $[32211]_{540}$ configurations. 
In Fig.~\ref{orbex} we show a schematic spectrum of the orbital 
excitations of the pentaquark up to $N=1$ quantum, which depends on 
the excitation energies, $\Delta_1$ and $\Delta_2$ 
\ba
\Delta_1 &=& E_{\rm orb}(1^-_{F_2})-E_{\rm orb}(0^+_{A_1}) ~,
\nonumber\\
\Delta_2 &=& E_{\rm orb}(1^-_{A_1})-E_{\rm orb}(0^+_{A_1}) ~. 
\ea

The energy of a given spin-flavour multiplet depends on the orbital 
excitation energies $\Delta_1$ and $\Delta_2$, and the coefficient $A$, 
while the terms proportional to $B$, $C$, $D$ and $E$ give the splitting 
inside the multiplet. The quantum numbers of the ground state depend on 
the relative size of $\Delta_1$ and $A$. Its parity is opposite to that 
of the orbital excitation due to the negative intrinsic parity of 
$q^4 \bar{q}$ configurations. 

For $\Delta_1 > 4A = 220$ MeV, the ground state pentaquark is associated 
with the orbital state with $L^p_t=0^+_{A_1}$ and the $^{2}\overline{10}$ 
anti-decuplet state of the $[42111]$ multiplet. In this 
case, the angular momentum and parity of the ground state pentaquark 
are $J^p=1/2^-$. Another possible identification of the observed 
$\Theta^{+}$ is provided by the $[42111]$ anti-decuplet state with 
$s=3/2$, in which case the ground state would have $J^p=3/2^-$. 
This would imply that, 
because of the positive value of the spin splitting coefficient $C$ in 
Eqs.~(\ref{grmass}) and (\ref{grpar}), there should be a another pentaquark 
state with $s=1/2$ and $J^p=1/2^-$ at an energy lower than the one observed. 
At the moment, there is no experimental evidence for such an exotic state  
for which reason this identification seems to be ruled out. 

For $\Delta_1 < 4A = 220$ MeV, the parity of the 
lowest pentaquark state would be positive, since the ground state now 
corresponds to the orbital excitation with $L^p_t=1^-_{F_2}$ and the 
$^{2}\overline{10}$ flavour anti-decuplet of the $[51111]$ multiplet.  
In the absence of spin-orbit splitting, we find in this case a ground state 
doublet with angular momentum and parity $J^p=1/2^+$, $3/2^+$. 
The calculation of Table~\ref{pentamass2} belongs to this class 
since $\Delta_1=0$. 

\section{Summary, conclusions and outlook}

In this work, we have constructed a  classification scheme of the 
pentaquark states in terms of $SU_{\rm sf}(6)$ spin-flavour multiplets, 
and their flavour and spin content in terms of $SU_{\rm f}(3)$ and 
$SU_{\rm s}(2)$ states. Exotic pentaquark states can be found only in the 
flavour anti-decuplets, 27-plets and 35-plets. Moreover, we have discussed 
the permutation symmetry properties of both the spin-flavour and orbital 
parts of the $qqqq$ subsystem. In order to obtain the total wave function, 
the spin-flavour part has been combined with the colour and  orbital 
contributions in such a way that the total pentaquark wave function  is a
colour singlet and is antisymmetric under the interchange of any of the 
four quarks. This classification scheme is general and complete, and may be 
helpful for both experimental, CQM and lattice QCD studies. In particular, 
the constructed basis for pentaquark states will enable to solve the 
eigenvalue problem for a definite dynamical model. This is valid 
not only for Constituent Quark Models, but also for diquark-diquark-antiquark 
approaches, for which the basis is a subset of the one we have constructed. 

As an application we have calculated the mass spectrum of exotic pentaquark 
states with the G\"ursey-Radicati mass formula which corresponds to the 
dynamical symmetry described by the chain of subgroups
\ba 
SU_{\rm sf}(6) \supset SU_{\rm f}(3) \otimes SU_{\rm s}(2) 
\supset SU_{\rm I}(2) \otimes U_{\rm Y}(1) \otimes SU_{\rm s}(2) ~, 
\label{chain}
\ea
and encodes the slightly broken symmetries of the strong interactions. 
In the assumption of a GR formula we have neglected the radial dependence 
of the $SU_{\rm sf}(6)$ spin-flavour quark interaction. 
The problems that arise from this kind of approximation have been discussed 
in the literature, nevertheless similar methods have been used 
in other studies of pentaquark states. In principle, the coefficients of
the GR applied to the $qqqq\overline{q}$ system should be obtained from a fit of the pentaquark
spectrum. This is however not possible at the moment, since we know at most two pentaquark
states. Therefore,under the assumption that the coefficients 
do not depend strongly on the structure of the quark system, we 
have calculated the pentaquark spectrum using the
coefficients taken from a prior study  of $qqq$ baryons \cite{gsv}, 
in order to get an idea of the general features of the spectrum. 
As a result we find that the lowest 
pentaquark is always an $^{2}\overline{10}$ anti-decuplet state with 
isospin $I=0$, in agreement with experimental evidence that  the 
$\Theta^+(1540)$ is an isosinglet.  
We also presented some preliminary results 
based on a generalized G\"ursey-Radicati mass formula which includes the 
invariant of the $SU_{\rm sf}(6)$ spin-flavour group, and a simple schematic 
model for the orbital excitations up to $N=1$ quantum. 

The angular momentum and parity of the ground state exotic pentaquark 
depends on the relative contribution of the orbital and spin-flavour parts 
of the mass operator. We find that if the splitting due to the 
$SU_{\rm sf}(6)$ spin-flavour term is large compared to that between the 
orbital states, the ground state pentaquark has positive parity 
\cite{stancu,helminen,glozman}, whereas for a relatively small 
spin-flavour splitting the parity of the lowest pentaquark state becomes 
negative. We notice that, in case of a positive parity ground state, 
there is a doublet with $J^p=1/2^+$, $3/2^+$ which, in the presence of a 
spin-orbit coupling term, would give rise to a pair of peaks. 
The effect of specific dynamical models on the pentaquark spectrum in 
general, and on the properties of its ground state in particular, using a space dependent
$SU(6)$ breaking interaction, will be presented in more detail in a separate publication
\cite{BGS}. 

The spectroscopy of exotic baryons will be a key testing ground for models 
of baryons and their structure. Especially the determination of the 
angular momentum and parity of the $\Theta^+(1540)$ will allow to 
distinguish between different approaches \cite{jennings}. 
Most theoretical studies predict a postive parity for the $\Theta^+$ 
\cite{diakonov,borisyuk,stancu,helminen,hosaka,glozman,diquark}, 
but there is also evidence for a negative parity from recent work on QCD 
sum rules \cite{sumrule} and lattice QCD \cite{lattice}. 
Other challenges include the search for excited exotics. 

\section*{Acknowledgments}

This work was supported in part by a research grant from CONACyT, 
M\'exico.

\clearpage

\begin{table}
\centering
\caption[]{Symmetry properties of three-quark states}
\vspace{15pt}
\label{qqq}
\begin{tabular}{cccccccc}
\hline
& & & & & & & \\
& & & & & \multicolumn{3}{c}{Dimension} \\
$D_3$ & $\sim$ & $S_3$ & Young tableau 
& Multiplicity & $SU(6)$ & $SU(3)$ & $SU(2)$ \\
& & & & & & & \\
\hline
& & & & & & & \\
$A_1$ & $\sim$ & $[3]$ & $\setlength{\unitlength}{1.0pt}
\begin{picture}(30,10)(0,0)
\thinlines
\put ( 0, 0) {\line (1,0){30}}
\put ( 0,10) {\line (1,0){30}}
\put ( 0, 0) {\line (0,1){10}}
\put (10, 0) {\line (0,1){10}}
\put (20, 0) {\line (0,1){10}}
\put (30, 0) {\line (0,1){10}}
\end{picture}$ & 1 & 56 & 10 & 4 \\
& & & & & & & \\
$E$ & $\sim$ & $[21]$ & $\setlength{\unitlength}{1.0pt}
\begin{picture}(20,20)(0,5)
\thinlines
\put ( 0, 0) {\line (1,0){10}}
\put ( 0,10) {\line (1,0){20}}
\put ( 0,20) {\line (1,0){20}}
\put ( 0, 0) {\line (0,1){20}}
\put (10, 0) {\line (0,1){20}}
\put (20,10) {\line (0,1){10}}
\end{picture}$ & 2 & 70 & 8 & 2 \\
& & & & & & & \\
$A_2$ & $\sim$ & $[111]$ & $\setlength{\unitlength}{1.0pt}
\begin{picture}(10,30)(0,10)
\thinlines
\put ( 0, 0) {\line (1,0){10}}
\put ( 0,10) {\line (1,0){10}}
\put ( 0,20) {\line (1,0){10}}
\put ( 0,30) {\line (1,0){10}}
\put ( 0, 0) {\line (0,1){30}}
\put (10, 0) {\line (0,1){30}}
\end{picture}$ & 1 & 20 & 1 & $-$ \\
& & & & & & & \\
& & & & & & & \\
\hline 
\end{tabular}
\end{table}

\begin{table}
\centering
\caption[]{Spin-flavour classification of $q^3$ states}
\vspace{15pt}
\label{sfqqq}
\begin{tabular}{llclcl}
\hline
& & & & & \\
$D_3$ & $SU_{\rm sf}(6)$ & $\supset$ & $SU_{\rm f}(3)$ 
& $\otimes$ & $SU_{\rm s}(2)$ \\
& & & & & \\
\hline
& & & & & \\
$A_1$ & $[3]_{56}$ & & $[3]_{10}$  & $\otimes$ & $[3]_{4}$ \\
                & & & $[21]_{8}$ & $\otimes$ & $[21]_{2}$ \\
& & & & & \\
$E$ & $[21]_{70}$ & & $[3]_{10}$  & $\otimes$ & $[21]_{2}$ \\
               & & & $[21]_{8}$ & $\otimes$ & $[3]_{4}$ \\
               & & & $[21]_{8}$ & $\otimes$ & $[21]_{2}$ \\
              & & & $[111]_{1}$ & $\otimes$ & $[21]_{2}$ \\
& & & & & \\
$A_2$ & $[111]_{20}$ & & $[21]_{8}$  & $\otimes$ & $[21]_{2}$ \\
              & & & $[111]_{1}$ & $\otimes$ & $[3]_{4}$ \\
& & & & & \\
\hline
\end{tabular}
\end{table}

\begin{table}
\centering
\caption[]{Symmetry properties of four-quark $SU(6)$ states}
\vspace{15pt}
\label{qqqq}
\begin{tabular}{cccccccc}
\hline
& & & & & & & \\
& & & & & \multicolumn{3}{c}{Dimension} \\
${\cal T}_d$ & $\sim$ & $S_4$ & Young tableau & Multiplicity 
& $SU(6)$ & $SU(3)$ & $SU(2)$ \\
& & & & & & & \\
\hline
& & & & & & & \\
$A_1$ & $\sim$ & $[4]$ & $\setlength{\unitlength}{1.0pt}
\begin{picture}(40,10)(0,0)
\thinlines
\put ( 0, 0) {\line (1,0){40}}
\put ( 0,10) {\line (1,0){40}}
\put ( 0, 0) {\line (0,1){10}}
\put (10, 0) {\line (0,1){10}}
\put (20, 0) {\line (0,1){10}}
\put (30, 0) {\line (0,1){10}}
\put (40, 0) {\line (0,1){10}}
\end{picture}$ & 1 & 126 & 15 & 5 \\
& & & & & & & \\
$F_2$ & $\sim$ & $[31]$ & $\setlength{\unitlength}{1.0pt}
\begin{picture}(30,20)(0,5)
\thinlines
\put ( 0, 0) {\line (1,0){10}}
\put ( 0,10) {\line (1,0){30}}
\put ( 0,20) {\line (1,0){30}}
\put ( 0, 0) {\line (0,1){20}}
\put (10, 0) {\line (0,1){20}}
\put (20,10) {\line (0,1){10}}
\put (30,10) {\line (0,1){10}}
\end{picture}$ & 3 & 210 & 15 & 3 \\
& & & & & & & \\
$E$ & $\sim$ & $[22]$ & $\setlength{\unitlength}{1.0pt}
\begin{picture}(20,20)(0,5)
\thinlines
\put ( 0, 0) {\line (1,0){20}}
\put ( 0,10) {\line (1,0){20}}
\put ( 0,20) {\line (1,0){20}}
\put ( 0, 0) {\line (0,1){20}}
\put (10, 0) {\line (0,1){20}}
\put (20, 0) {\line (0,1){20}}
\end{picture}$ & 2 & 105 & 6 & 1 \\
& & & & & & & \\
$F_1$ & $\sim$ & $[211]$ & $\setlength{\unitlength}{1.0pt}
\begin{picture}(20,30)(0,10)
\thinlines
\put ( 0, 0) {\line (1,0){10}}
\put ( 0,10) {\line (1,0){10}}
\put ( 0,20) {\line (1,0){20}}
\put ( 0,30) {\line (1,0){20}}
\put ( 0, 0) {\line (0,1){30}}
\put (10, 0) {\line (0,1){30}}
\put (20,20) {\line (0,1){10}}
\end{picture}$ & 3 & 105 & 3 & $-$ \\
& & & & & & & \\
$A_2$ & $\sim$ & $[1111]$ & $\setlength{\unitlength}{1.0pt}
\begin{picture}(10,40)(0,15)
\thinlines
\put ( 0, 0) {\line (1,0){10}}
\put ( 0,10) {\line (1,0){10}}
\put ( 0,20) {\line (1,0){10}}
\put ( 0,30) {\line (1,0){10}}
\put ( 0,40) {\line (1,0){10}}
\put ( 0, 0) {\line (0,1){40}}
\put (10, 0) {\line (0,1){40}}
\end{picture}$ & 1 & 15 & $-$ & $-$ \\
& & & & & & & \\
& & & & & & & \\
\hline 
\end{tabular}
\end{table}

\begin{table}
\centering
\caption[]{Spin-flavour decomposition of $q^4$ states}
\vspace{15pt}
\label{sfqqqq}
\begin{tabular}{cllclcl}
\hline
& & & & & & \\
$D_3$ & ${\cal T}_d$ & $SU_{\rm sf}(6)$ & $\supset$ & $SU_{\rm f}(3)$ 
& $\otimes$ & $SU_{\rm s}(2)$ \\
& & & & & & \\
\hline
& & & & & & \\
$A_1$ & $A_1$ & $[4]_{126}$ & & $[4]_{15}$  & $\otimes$ & $[4]_{5}$ \\
                    & & & & $[31]_{15}$ & $\otimes$ & $[31]_{3}$ \\
                    & & & & $[22]_{6}$  & $\otimes$ & $[22]_{1}$ \\
& & & & & & \\
$A_1+E$ & $F_2$ & $[31]_{210}$ & & $[4]_{15}$  & $\otimes$ & $[31]_{3}$ \\
              & & & & $[31]_{15}$ & $\otimes$ & $[4]_{5}$ \\
              & & & & $[31]_{15}$ & $\otimes$ & $[31]_{3}$ \\
              & & & & $[31]_{15}$ & $\otimes$ & $[22]_{1}$ \\
              & & & & $[22]_{6}$  & $\otimes$ & $[31]_{3}$ \\
              & & & & $[211]_{3}$ & $\otimes$ & $[22]_{1}$ \\
              & & & & $[211]_{3}$ & $\otimes$ & $[31]_{3}$ \\
& & & & & & \\
$E$ & $E$ & $[22]_{105}$ & & $[4]_{15}$  & $\otimes$ & $[22]_{1}$ \\
              & & & & $[31]_{15}$ & $\otimes$ & $[31]_{3}$ \\
              & & & & $[22]_{6}$  & $\otimes$ & $[4]_{5}$ \\
              & & & & $[22]_{6}$  & $\otimes$ & $[22]_{1}$ \\
              & & & & $[211]_{3}$ & $\otimes$ & $[31]_{3}$ \\
& & & & & & \\
$E+A_2$ & $F_1$ & $[211]_{105}$ & & $[31]_{15}$ & $\otimes$ & $[31]_{3}$ \\
               & & & & $[31]_{15}$ & $\otimes$ & $[22]_{1}$ \\
               & & & & $[22]_{6}$  & $\otimes$ & $[31]_{3}$ \\
               & & & & $[211]_{3}$ & $\otimes$ & $[4]_{5}$ \\
               & & & & $[211]_{3}$ & $\otimes$ & $[31]_{3}$ \\
               & & & & $[211]_{3}$ & $\otimes$ & $[22]_{1}$ \\
& & & & & & \\
$A_2$ & $A_2$ & $[1111]_{15}$ & & $[22]_{6}$  & $\otimes$ & $[22]_{1}$ \\
              & & & & $[211]_{3}$ & $\otimes$ & $[31]_{3}$ \\
& & & & & & \\
\hline
\end{tabular}
\end{table}

\begin{table}
\centering
\caption[]{Allowed spin, flavour and spin-flavour pentaquark states}
\vspace{15pt}
\label{states}
$\begin{array}{cccc}
\hline
& & & \\
& qqqq\bar{q} & \mbox{Dimension} & S_4 \sim {\cal T}_d \\
& & & \\
\hline
& & & \\
\mbox{spin} & [5] & 6 & A_1 \\
& [41] & 4 & A_1, F_2 \\
& [32] & 2 & F_2, E   \\
& & & \\
\hline
& & & \\
\mbox{flavour} & [51] & \mbox{35-plet} & A_1 \\
& [42]  & \mbox{27-plet} & F_2 \\
& [33]  & \mbox{antidecuplet} & E \\
& [411] & \mbox{decuplet} & A_1, F_2 \\
& [321] & \mbox{octet} & F_2, E, F_1 \\
& [222] & \mbox{singlet} & F_1 \\
& & & \\
\hline
& & & \\
\mbox{spin-flavour} 
& [51111]  &  700 & A_1 \\
& [411111] &   56 & A_1, F_2 \\
& [42111]  & 1134 & F_2 \\
& [321111] &   70 & F_2, E, F_1 \\
& [33111]  &  560 & E \\
& [32211]  &  540 & F_1 \\
& [222111] &   20 & F_1, A_2 \\
& [22221]  &   70 & A_2 \\
& & & \\
\hline
\end{array}$
\end{table}

\clearpage

\begin{table}
\centering
\caption[]{Spin-flavour classification of $q^4 \bar{q}$ states. 
The ${\cal T}_d$ labels refer to the $q^4$ subsystem.}
\vspace{15pt}
\label{pentaquark}
\begin{tabular}{clclcl}
\hline
& & & & & \\
${\cal T}_d$ & $SU_{\rm sf}(6)$ & $\supset$ & $SU_{\rm f}(3)$ 
& $\otimes$ & $SU_{\rm s}(2)$ \\
& & & & & \\
\hline
& & & & & \\
$A_1$ & $[51111]_{700}$ 
  & & $[51]_{35}$  & $\otimes$ & $[5]_{6}$ \\
& & & $[51]_{35}$  & $\otimes$ & $[41]_{4}$ \\
& & & $[42]_{27}$  & $\otimes$ & $[41]_{4}$ \\
& & & $[42]_{27}$  & $\otimes$ & $[32]_{2}$ \\
& & & $[33]_{10}$  & $\otimes$ & $[32]_{2}$ \\
& & & $[411]_{10}$ & $\otimes$ & $[5]_{6}$ \\
& & & $[411]_{10}$ & $\otimes$ & $[41]_{4}$ \\
& & & $[411]_{10}$ & $\otimes$ & $[32]_{2}$ \\
& & & $[321]_{8}$  & $\otimes$ & $[41]_{4}$ \\
& & & $[321]_{8}$  & $\otimes$ & $[32]_{2}$ \\
& & & & & \\
$A_1+F_2$ & $[411111]_{56}$ 
  & & $[411]_{10}$ & $\otimes$ & $[41]_{4}$ \\
& & & $[321]_{8}$  & $\otimes$ & $[32]_{2}$ \\
& & & & & \\
$F_2$ & $[42111]_{1134}$ 
  & & $[51]_{35}$  & $\otimes$ & $[41]_{4}$ \\
& & & $[51]_{35}$  & $\otimes$ & $[32]_{2}$ \\
& & & $[42]_{27}$  & $\otimes$ & $[5]_{6}$ \\
& & & $2([42]_{27}$  & $\otimes$ & $[41]_{4}$) \\
& & & $2([42]_{27}$  & $\otimes$ & $[32]_{2}$) \\
& & & $[33]_{10}$  & $\otimes$ & $[41]_{4}$ \\
& & & $[33]_{10}$  & $\otimes$ & $[32]_{2}$ \\
& & & $[411]_{10}$ & $\otimes$ & $[5]_{6}$ \\
& & & $2([411]_{10}$ & $\otimes$ & $[41]_{4}$) \\
& & & $2([411]_{10}$ & $\otimes$ & $[32]_{2}$) \\
& & & $[321]_{8}$  & $\otimes$ & $[5]_{6}$ \\
& & & $3([321]_{8}$  & $\otimes$ & $[41]_{4}$) \\
& & & $3([321]_{8}$  & $\otimes$ & $[32]_{2}$) \\
& & & $[222]_{1}$  & $\otimes$ & $[41]_{4}$ \\
& & & $[222]_{1}$  & $\otimes$ & $[32]_{2}$ \\
& & & & & \\
$F_2+E+F_1$ & $[321111]_{70}$ 
  & & $[411]_{10}$ & $\otimes$ & $[32]_{2}$ \\
& & & $[321]_{8}$  & $\otimes$ & $[41]_{4}$ \\
& & & $[321]_{8}$  & $\otimes$ & $[32]_{2}$ \\
& & & $[222]_{1}$  & $\otimes$ & $[32]_{2}$ \\
& & & & & \\
\hline
\end{tabular}
\end{table}

\addtocounter{table}{-1}

\begin{table}
\centering
\caption[]{Continued}
\vspace{15pt}
\begin{tabular}{clclcl}
\hline
& & & & & \\
${\cal T}_d$ & $SU_{\rm sf}(6)$ & $\supset$ & $SU_{\rm f}(3)$ 
& $\otimes$ & $SU_{\rm s}(2)$ \\
& & & & & \\
\hline
& & & & & \\
$E$ & $[33111]_{560}$ 
  & & $[51]_{35}$  & $\otimes$ & $[32]_{2}$ \\
& & & $[42]_{27}$  & $\otimes$ & $[41]_{4}$ \\
& & & $[42]_{27}$  & $\otimes$ & $[32]_{2}$ \\
& & & $[33]_{10}$  & $\otimes$ & $[5]_{6}$ \\
& & & $[33]_{10}$  & $\otimes$ & $[41]_{4}$ \\
& & & $[33]_{10}$  & $\otimes$ & $[32]_{2}$ \\
& & & $[411]_{10}$ & $\otimes$ & $[41]_{4}$ \\
& & & $[411]_{10}$ & $\otimes$ & $[32]_{2}$ \\
& & & $[321]_{8}$  & $\otimes$ & $[5]_{6}$ \\
& & & $2([321]_{8}$  & $\otimes$ & $[41]_{4}$) \\
& & & $2([321]_{8}$  & $\otimes$ & $[32]_{2}$) \\
& & & $[222]_{1}$  & $\otimes$ & $[41]_{4}$ \\
& & & & & \\
$F_1$ & $[32211]_{540}$ 
  & & $[42]_{27}$  & $\otimes$ & $[41]_{4}$ \\
& & & $2([42]_{27}$  & $\otimes$ & $[32]_{2}$) \\
& & & $[33]_{10}$  & $\otimes$ & $[41]_{4}$ \\
& & & $[33]_{10}$  & $\otimes$ & $[32]_{2}$ \\
& & & $[411]_{10}$ & $\otimes$ & $[41]_{4}$ \\
& & & $[411]_{10}$ & $\otimes$ & $[32]_{2}$ \\
& & & $[321]_{8}$  & $\otimes$ & $[5]_{6}$ \\
& & & $3([321]_{8}$  & $\otimes$ & $[41]_{4}$) \\
& & & $3([321]_{8}$  & $\otimes$ & $[32]_{2}$) \\
& & & $[222]_{1}$  & $\otimes$ & $[5]_{6}$ \\
& & & $[222]_{1}$  & $\otimes$ & $[41]_{4}$ \\
& & & $[222]_{1}$  & $\otimes$ & $[32]_{2}$ \\
& & & & & \\
$F_1+A_2$ & $[222111]_{20}$ 
  & & $[321]_{8}$  & $\otimes$ & $[32]_{2}$ \\
& & & $[222]_{1}$  & $\otimes$ & $[41]_{4}$ \\
& & & & & \\
$A_2$ & $[22221]_{70}$ 
  & & $[33]_{10}$ & $\otimes$ & $[32]_{2}$ \\
& & & $[321]_{8}$  & $\otimes$ & $[41]_{4}$ \\
& & & $[321]_{8}$  & $\otimes$ & $[32]_{2}$ \\
& & & $[222]_{1}$  & $\otimes$ & $[32]_{2}$ \\
& & & & & \\
\hline
\end{tabular}
\end{table}

\begin{table}
\centering
\caption[]{$q^4 \bar{q}$ pentaquark states with exotic quantum numbers. 
The electric charge is $Q=I_3+Y/2$. Notation as in \protect\cite{notation}.}
\vspace{15pt}
\label{exotic}
\begin{tabular}{crcccccc}
\hline
& & & & & & & \\
$SU_{\rm f}(3)$ & $Y$ & $I$ & $Q$ & Flavour States & Notation \\
& & & & & & & \\
\hline
& & & & & & & \\
$[33]_{10}$ &  2 &  0  &    1 & $dduu\bar{s}$ & $\Theta$ \\
            & -1 & 3/2 & -2,1 & $ddss\bar{u},uuss\bar{d}$ & $\Xi_{3/2}$ \\
& & & & & & & \\
\hline
& & & & & & & \\
$[42]_{27}$ &  2 &  1  & 0,1,2 & $dddu\bar{s}$, $dduu\bar{s}$, $duuu\bar{s}$ 
& $\Theta_1$ \\
            &  0 &  2  & -2,2  & $ddds\bar{u}$, $uuus\bar{d}$ & $\Sigma_2$ \\
            & -1 & 3/2 & -2,1  & $ddss\bar{u}$, $uuss\bar{d}$ & $\Xi_{3/2}$ \\
            & -2 &  1  & -2,0  & $dsss\bar{u}$, $usss\bar{d}$ & $\Omega_1$ \\
& & & & & & & \\
\hline
& & & & & & & \\
$[51]_{35}$ &  2 &  2  & -1,0,1,2,3 & $dddd\bar{s}$, $dddu\bar{s}$, 
$dduu\bar{s}$, $duuu\bar{s}$, $uuuu\bar{s}$ & $\Theta_2$ \\
            &  1 & 5/2 & -2, 3 & $dddd\bar{u}$, $uuuu\bar{d}$ & $\Delta_{5/2}$ \\
            &  0 &  2  & -2, 2 & $ddds\bar{u}$, $uuus\bar{d}$ & $\Sigma_2$ \\
            & -1 & 3/2 & -2, 1 & $ddss\bar{u}$, $uuss\bar{d}$ & $\Xi_{3/2}$ \\
            & -2 &  1  & -2, 0 & $dsss\bar{u}$, $usss\bar{d}$ & $\Omega_1$ \\
            & -3 & 1/2 & -2,-1 & $ssss\bar{u}$, $ssss\bar{d}$ & $\Phi$ \\
& & & & & & & \\
\hline
\end{tabular}
\end{table}


\begin{table}
\centering
\caption[]{Decomposition of the orbital-spin-flavour wave function 
with $F_2$ symmetry into orbital and spin-flavour parts. In the last 
column the pentaquark configurations that contain exotic states 
are shown.}
\vspace{15pt}
\label{orbital}
\begin{tabular}{ccc}
\hline
& & \\
Orbital & Spin-Flavour & $q^4\bar{q}$ Configuration \\
Symmetry & Symmetry & with Exotic States \\
& & \\
\hline
& & \\
$A_1$ & $F_2$ & $[42111]$ \\
& & \\
$F_2$ & $A_1$ & $[51111]$ \\
      & $F_2$ & $[42111]$ \\
      & $E$   & $[33111]$ \\
      & $F_1$ & $[32211]$ \\
& & \\
$E$   & $F_2$ & $[42111]$ \\
      & $F_1$ & $[32211]$ \\
& & \\
$F_1$ & $A_2$ & $[22221]$ \\
      & $F_2$ & $[42111]$ \\
      & $E$   & $[33111]$ \\
      & $F_1$ & $[32211]$ \\
& & \\
$A_2$ & $F_1$ & $[32211]$ \\
& & \\
\hline
\end{tabular}
\end{table}

\begin{table}
\centering
\caption[]{Eigenvalues of the $C_{2SU_{\rm sf}(6)}$ and 
$C_{2SU_{\rm f}(3)}$ Casimir operators}
\vspace{15pt}
\label{casimir}
\begin{tabular}{lclc}
\hline
& & & \\
spin-flavour & $C_{2SU_{\rm sf}(6)}$ & flavour & $C_{2SU_{\rm f}(3)}$ \\
& & & \\
\hline
& & & \\
$[51111]_{700}$  & $81/4$ & $[51]_{35}$  & 12 \\
$[411111]_{56}$  & $45/4$ & $[42]_{27}$  &  8 \\
$[42111]_{1134}$ & $65/4$ & $[33]_{10}$  &  6 \\
$[321111]_{70}$  & $33/4$ & $[411]_{10}$ &  6 \\
$[33111]_{560}$  & $57/4$ & $[321]_{8}$  &  3 \\
$[32211]_{540}$  & $49/4$ & $[222]_{1}$  &  0 \\
$[222111]_{20}$  & $21/4$ & & \\
$[22221]_{70}$   & $33/4$ & & \\
& & & \\
\hline
\end{tabular}
\end{table}

\clearpage

\begin{table}[ht]
\centering
\caption[]{Mass splittings of exotic pentaquark states within a 
$SU_{\rm sf}(6)$ multiplet calculated using Eq.~(\ref{grmass}) with the 
parameters of Eq.~(\ref{grpar}). The pentaquark ground state is 
normalized to the observed mass of the $\Theta^+(1540)$ resonance. 
The orbital excitations are taken to be degenerate. The states are labeled 
by their spin $s$, hypercharge $Y$, isospin $I$, spin-flavour multiplet 
$[f]$ and orbital excitation $L^p_t$. The notation is the 
same as in Table~\protect\ref{exotic}.}
\vspace{15pt}
\label{pentamass1}
\begin{tabular}{ccrcccccc}
\hline
& & & & & & & & \\
& & & & & \multicolumn{4}{c}{Mass (MeV)} \\
$SU_{\rm f}(3)$ & $s$ & $Y$ & $I$ & Notation 
& $[51111]$ & $[42111]$ & $[33111]$ & $[32211]$ \\
& & & & & $1^-_{F_2}$ & $0^+_{A_1}$, $1^-_{A_1,F_2}$ & $1^-_{F_2}$ 
& $1^-_{F_2}$ \\
& & & & & & & & \\
\hline
& & & & & & & & \\
$[33]_{10}$ & 1/2 &  2 &  0  & $\Theta$  & {\bf 1540} & 
{\bf 1540} & {\bf 1540} & {\bf 1540} \\
            &     & -1 & 3/2 & $\Xi_{3/2}$ & 2305 & 2305 & 2305 & 2305 \\
& & & & & & & & \\
$[33]_{10}$ & 3/2 &  2 &  0  & $\Theta$  & & 1655 & 1655 & 1655 \\
            &     & -1 & 3/2 & $\Xi_{3/2}$ & & 2420 & 2420 & 2420 \\
& & & & & & & & \\
$[33]_{10}$ & 5/2 &  2 &  0  & $\Theta$  & & & 1846 & \\
            &     & -1 & 3/2 & $\Xi_{3/2}$ & & & 2612 & \\
& & & & & & & & \\
\hline
& & & & & & & & \\
$[42]_{27}$ & 1/2 &  2 &  1  & $\Theta_1$  & 1659 & 1659 & 1659 & 1659 \\
            &     &  0 &  2  & $\Sigma_2$  & 2247 & 2247 & 2247 & 2247 \\
            &     & -1 & 3/2 & $\Xi_{3/2}$ & 2348 & 2348 & 2348 & 2348 \\
            &     & -2 &  1  & $\Omega_1$  & 2449 & 2449 & 2449 & 2449 \\
& & & & & & & & \\
$[42]_{27}$ & 3/2 &  2 &  1  & $\Theta_1$  & 1774 & 1774 & 1774 & 1774 \\
            &     &  0 &  2  & $\Sigma_2$  & 2361 & 2361 & 2361 & 2361 \\
            &     & -1 & 3/2 & $\Xi_{3/2}$ & 2461 & 2461 & 2461 & 2461 \\
            &     & -2 &  1  & $\Omega_1$  & 2564 & 2564 & 2564 & 2564 \\
& & & & & & & & \\
$[42]_{27}$ & 5/2 &  2 &  1  & $\Theta_1$  & & 1966 & & \\
            &     &  0 &  2  & $\Sigma_2$  & & 2553 & & \\
            &     & -1 & 3/2 & $\Xi_{3/2}$ & & 2654 & & \\
            &     & -2 &  1  & $\Omega_1$  & & 2755 & & \\
& & & & & & & & \\
\hline
\end{tabular}
\end{table}

\addtocounter{table}{-1}

\begin{table}
\centering
\caption[]{Continued}
\vspace{15pt}
\begin{tabular}{ccrcccccc}
\hline
& & & & & & & & \\
& & & & & \multicolumn{4}{c}{Mass (MeV)} \\
$SU_{\rm f}(3)$ & $s$ & $Y$ & $I$ & Notation 
& $[51111]$ & $[42111]$ & $[33111]$ & $[32211]$ \\
& & & & & $1^-_{F_2}$ & $0^+_{A_1}$, $1^-_{A_1,F_2}$ & $1^-_{F_2}$ 
& $1^-_{F_2}$ \\
& & & & & & & & \\
\hline
& & & & & & & & \\
$[51]_{35}$ & 1/2 &  2 &  2  & $\Theta_2$     & & 1898 & 1898 & \\
            &     &  1 & 5/2 & $\Delta_{5/2}$ & & 2230 & 2230 & \\
            &     &  0 &  2  & $\Sigma_2$     & & 2331 & 2331 & \\
            &     & -1 & 3/2 & $\Xi_{3/2}$    & & 2432 & 2432 & \\
            &     & -2 &  1  & $\Omega_1$     & & 2533 & 2533 & \\
            &     & -3 & 1/2 & $\Phi$         & & 2634 & 2634 & \\
& & & & & & & & \\
$[51]_{35}$ & 3/2 &  2 &  2  & $\Theta_2$     & 2013 & 2013 & & \\
            &     &  1 & 5/2 & $\Delta_{5/2}$ & 2345 & 2345 & & \\
            &     &  0 &  2  & $\Sigma_2$     & 2446 & 2446 & & \\
            &     & -1 & 3/2 & $\Xi_{3/2}$    & 2547 & 2547 & & \\
            &     & -2 &  1  & $\Omega_1$     & 2648 & 2648 & & \\
            &     & -3 & 1/2 & $\Phi$         & 2749 & 2749 & & \\
& & & & & & & & \\
$[51]_{35}$ & 5/2 &  2 &  2  & $\Theta_2$     & 2205 & & & \\
            &     &  1 & 5/2 & $\Delta_{5/2}$ & 2537 & & & \\
            &     &  0 &  2  & $\Sigma_2$     & 2638 & & & \\
            &     & -1 & 3/2 & $\Xi_{3/2}$    & 2739 & & & \\
            &     & -2 &  1  & $\Omega_1$     & 2840 & & & \\
            &     & -3 & 1/2 & $\Phi$         & 2941 & & & \\
& & & & & & & & \\
\hline
\end{tabular}
\end{table}

\begin{table}[ht]
\centering
\caption[]{Spin-flavour contribution to the masses of exotic pentaquark 
states calculated using Eq.~(\ref{grmass}) with the parameters of 
Eq.~(\ref{grpar}) and $A=55.1$ MeV. The pentaquark ground state is 
normalized to the observed mass of the $\Theta^+(1540)$ resonance. 
The notation and the labeling of the states is the same as in 
Table~\protect\ref{pentamass1}. 
The orbital excitations are taken to be degenerate.}
\vspace{15pt}
\label{pentamass2}
\begin{tabular}{ccrcccccc}
\hline
& & & & & & & & \\
& & & & & \multicolumn{4}{c}{Mass (MeV)} \\
$SU_{\rm f}(3)$ & $s$ & $Y$ & $I$ & Notation 
& $[51111]$ & $[42111]$ & $[33111]$ & $[32211]$ \\
& & & & & $1^-_{F_2}$ & $0^+_{A_1}$, $1^-_{A_1,F_2}$ & $1^-_{F_2}$ 
& $1^-_{F_2}$ \\
& & & & & & & & \\
\hline
& & & & & & & & \\
$[33]_{10}$ & 1/2 &  2 &  0  & $\Theta$  & 1320 & {\bf 1540} & 1650 & 1760 \\
            &     & -1 & 3/2 & $\Xi_{3/2}$ & 2085 & 2305 & 2415 & 2526 \\
& & & & & & & & \\
$[33]_{10}$ & 3/2 &  2 &  0  & $\Theta$    & & 1655 & 1765 & 1875 \\
            &     & -1 & 3/2 & $\Xi_{3/2}$ & & 2420 & 2530 & 2640 \\
& & & & & & & & \\
$[33]_{10}$ & 5/2 &  2 &  0  & $\Theta$    & & & 1957 & \\
            &     & -1 & 3/2 & $\Xi_{3/2}$ & & & 2722 & \\
& & & & & & & & \\
\hline
& & & & & & & & \\
$[42]_{27}$ & 1/2 &  2 &  1  & $\Theta_1$  & 1439 & 1659 & 1770 & 1880 \\
            &     &  0 &  2  & $\Sigma_2$  & 2026 & 2247 & 2357 & 2467 \\
            &     & -1 & 3/2 & $\Xi_{3/2}$ & 2127 & 2348 & 2458 & 2568 \\
            &     & -2 &  1  & $\Omega_1$  & 2228 & 2449 & 2559 & 2669 \\
& & & & & & & & \\
$[42]_{27}$ & 3/2 &  2 &  1  & $\Theta_1$  & 1554 & 1774 & 1885 & 1995 \\
            &     &  0 &  2  & $\Sigma_2$  & 2141 & 2361 & 2472 & 2582 \\
            &     & -1 & 3/2 & $\Xi_{3/2}$ & 2242 & 2462 & 2573 & 2683 \\
            &     & -2 &  1  & $\Omega_1$  & 2343 & 2564 & 2674 & 2784 \\
& & & & & & & & \\
$[42]_{27}$ & 5/2 &  2 &  1  & $\Theta_1$  & & 1966 & & \\
            &     &  0 &  2  & $\Sigma_2$  & & 2553 & & \\
            &     & -1 & 3/2 & $\Xi_{3/2}$ & & 2654 & & \\
            &     & -2 &  1  & $\Omega_1$  & & 2755 & & \\
& & & & & & & & \\
\hline
\end{tabular}
\end{table}

\addtocounter{table}{-1}


\begin{table}
\centering
\caption[]{Continued}
\vspace{15pt}
\begin{tabular}{ccrcccccc}
\hline
& & & & & & & & \\
& & & & & \multicolumn{4}{c}{Mass (MeV)} \\
$SU_{\rm f}(3)$ & $s$ & $Y$ & $I$ & Notation 
& $[51111]$ & $[42111]$ & $[33111]$ & $[32211]$ \\
& & & & & $1^-_{F_2}$ & $0^+_{A_1}$, $1^-_{A_1,F_2}$ & $1^-_{F_2}$ 
& $1^-_{F_2}$ \\
& & & & & & & & \\
\hline
& & & & & & & & \\
$[51]_{35}$ & 1/2 &  2 &  2  & $\Theta_2$     & & 1898 & 2008 & \\
            &     &  1 & 5/2 & $\Delta_{5/2}$ & & 2230 & 2340 & \\
            &     &  0 &  2  & $\Sigma_2$     & & 2331 & 2442 & \\
            &     & -1 & 3/2 & $\Xi_{3/2}$    & & 2432 & 2543 & \\
            &     & -2 &  1  & $\Omega_1$     & & 2533 & 2644 & \\
            &     & -3 & 1/2 & $\Phi$         & & 2634 & 2745 & \\
& & & & & & & & \\
$[51]_{35}$ & 3/2 &  2 &  2  & $\Theta_2$     & 1793 & 2013 & & \\
            &     &  1 & 5/2 & $\Delta_{5/2}$ & 2125 & 2345 & & \\
            &     &  0 &  2  & $\Sigma_2$     & 2226 & 2446 & & \\
            &     & -1 & 3/2 & $\Xi_{3/2}$    & 2327 & 2547 & & \\
            &     & -2 &  1  & $\Omega_1$     & 2428 & 2648 & & \\
            &     & -3 & 1/2 & $\Phi$         & 2529 & 2749 & & \\
& & & & & & & & \\
$[51]_{35}$ & 5/2 &  2 &  2  & $\Theta_2$     & 1984 & & & \\
            &     &  1 & 5/2 & $\Delta_{5/2}$ & 2316 & & & \\
            &     &  0 &  2  & $\Sigma_2$     & 2417 & & & \\
            &     & -1 & 3/2 & $\Xi_{3/2}$    & 2518 & & & \\
            &     & -2 &  1  & $\Omega_1$     & 2619 & & & \\
            &     & -3 & 1/2 & $\Phi$         & 2720 & & & \\
& & & & & & & & \\
\hline
\end{tabular}
\end{table}

\clearpage

\begin{figure}
\centering
\setlength{\unitlength}{0.6pt}
\begin{picture}(350,200)(75,75)
\thinlines
\put(200,200) {\line(1,0){100}}
\put(150,150) {\line(1,0){200}}
\put(100,100) {\line(1,0){300}}
\put(100,100) {\line(1,1){150}}
\put(200,100) {\line(1,1){100}}
\put(300,100) {\line(1,1){ 50}}
\put(200,100) {\line(-1,1){ 50}}
\put(300,100) {\line(-1,1){100}}
\put(400,100) {\line(-1,1){150}}
\multiput(100,100)(300,0){2}{\circle*{10}}
\put(250,250){\circle*{10}}
\put(230,265){$dduu\bar{s}$}
\put( 80, 75){$ddss\bar{u}$}
\put(380, 75){$uuss\bar{d}$}
\end{picture}
\caption[]{$SU(3)$ flavour multiplet $[33]_{10}$ with $E$ symmetry. 
The isospin-hypercharge multiplets are $(I,Y)=(0,2)$, $(1/2,1)$, $(1,0)$ 
and $(3/2,-1)$. Exotic states are indicated with $\bullet$.}
\label{flavour33e}
\end{figure}
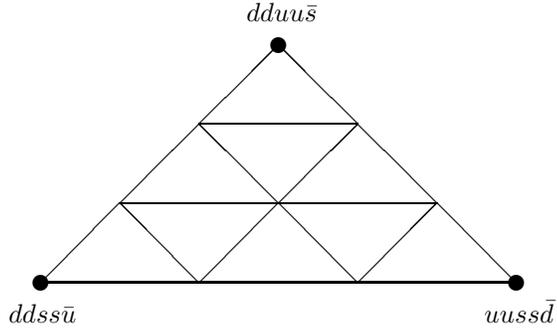

\begin{figure}
\centering
\setlength{\unitlength}{0.6pt}
\begin{picture}(450,250)(25,25)
\thinlines
\put(150,250) {\line(1,0){200}}
\put(100,200) {\line(1,0){300}}
\put( 50,150) {\line(1,0){400}}
\put(100,100) {\line(1,0){300}}
\put(150, 50) {\line(1,0){200}}
\put( 50,150) {\line(1,1){100}}
\put(100,100) {\line(1,1){150}}
\put(150, 50) {\line(1,1){200}}
\put(250, 50) {\line(1,1){150}}
\put(350, 50) {\line(1,1){100}}
\put(150, 50) {\line(-1,1){100}}
\put(250, 50) {\line(-1,1){150}}
\put(350, 50) {\line(-1,1){200}}
\put(400,100) {\line(-1,1){150}}
\put(450,150) {\line(-1,1){100}}

\multiput(150,250)(100,0){3}{\circle*{10}}
\multiput( 50,150)(400,0){2}{\circle*{10}}
\multiput(100,100)(300,0){2}{\circle*{10}}
\multiput(150, 50)(200,0){2}{\circle*{10}}

\multiput(200,200)(100,0){2}{\circle{10}}
\multiput(150,150)(100,0){3}{\circle{10}}
\multiput(200,100)(100,0){2}{\circle{10}}
\multiput(250,150)(100,0){1}{\circle{15}}

\put(125,265){$dddu\bar{s}$}
\put(225,265){$dduu\bar{s}$}
\put(325,265){$duuu\bar{s}$}
\put(-10,145){$ddds\bar{u}$}
\put(465,145){$uuus\bar{d}$}
\put( 40, 95){$ddss\bar{u}$}
\put(415, 95){$uuss\bar{d}$}
\put(125, 25){$dsss\bar{u}$}
\put(325, 25){$usss\bar{d}$}
\end{picture}
\caption[]{$SU(3)$ flavour multiplet $[42]_{27}$ with $F_2$ symmetry. 
The isospin-hypercharge multiplets are $(I,Y)=(1,2)$, $(3/2,1)$, $(1/2,1)$, 
$(2,0)$, $(1,0)$, $(0,0)$, $(3/2,-1)$, $(1/2,-1)$ and $(1,-2)$. 
Exotic states are indicated with $\bullet$.}
\label{flavour42f2}
\end{figure}
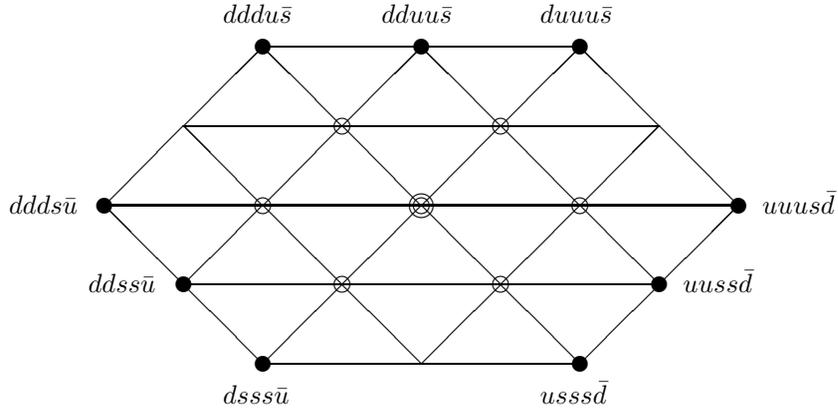

\begin{figure}
\centering
\setlength{\unitlength}{0.6pt}
\begin{picture}(550,300)(-25,-25)
\thinlines
\put( 50,250) {\line( 1,0){400}}
\put(  0,200) {\line( 1,0){500}}
\put( 50,150) {\line( 1,0){400}}
\put(100,100) {\line( 1,0){300}}
\put(150, 50) {\line( 1,0){200}}
\put(200,  0) {\line( 1,0){100}}
\put(  0,200) {\line( 1,1){ 50}}
\put( 50,150) {\line( 1,1){100}}
\put(100,100) {\line( 1,1){150}}
\put(150, 50) {\line( 1,1){200}}
\put(200,  0) {\line( 1,1){250}}
\put(300,  0) {\line( 1,1){200}}
\put(200,  0) {\line(-1,1){200}}
\put(300,  0) {\line(-1,1){250}}
\put(350, 50) {\line(-1,1){200}}
\put(400,100) {\line(-1,1){150}}
\put(450,150) {\line(-1,1){100}}
\put(500,200) {\line(-1,1){ 50}}

\multiput( 50,250)(100,0){5}{\circle*{10}}
\multiput(  0,200)(500,0){2}{\circle*{10}}
\multiput( 50,150)(400,0){2}{\circle*{10}}
\multiput(100,100)(300,0){2}{\circle*{10}}
\multiput(150, 50)(200,0){2}{\circle*{10}}
\multiput(200,  0)(100,0){2}{\circle*{10}}

\multiput(100,200)(100,0){4}{\circle{10}}
\multiput(150,150)(100,0){3}{\circle{10}}
\multiput(200,100)(100,0){2}{\circle{10}}
\multiput(250, 50)(100,0){1}{\circle{10}}

\put( 30,265){$dddd\bar{s}$}
\put(130,265){$dddu\bar{s}$}
\put(230,265){$dduu\bar{s}$}
\put(330,265){$duuu\bar{s}$}
\put(430,265){$uuuu\bar{s}$}
\put(-40,215){$dddd\bar{u}$}
\put(-10,145){$ddds\bar{u}$}
\put( 40, 95){$ddss\bar{u}$}
\put( 90, 45){$dsss\bar{u}$}
\put(180,-25){$ssss\bar{u}$}
\put(495,215){$uuuu\bar{d}$}
\put(465,145){$uuus\bar{d}$}
\put(415, 95){$uuss\bar{d}$}
\put(365, 45){$usss\bar{d}$}
\put(280,-25){$ssss\bar{d}$}
\end{picture}
\caption[]{$SU(3)$ flavour multiplet $[51]_{35}$ with $A_1$ symmetry. 
The isospin-hypercharge multiplets are $(I,Y)=(2,2)$, $(5/2,1)$, $(3/2,1)$, 
$(2,0)$, $(1,0)$, $(3/2,-1)$, $(1/2,-1)$, $(1,-2)$, $(0,-2)$ and 
$(1/2,-3)$. Exotic states are indicated with $\bullet$.}
\label{flavour51a1}
\end{figure}
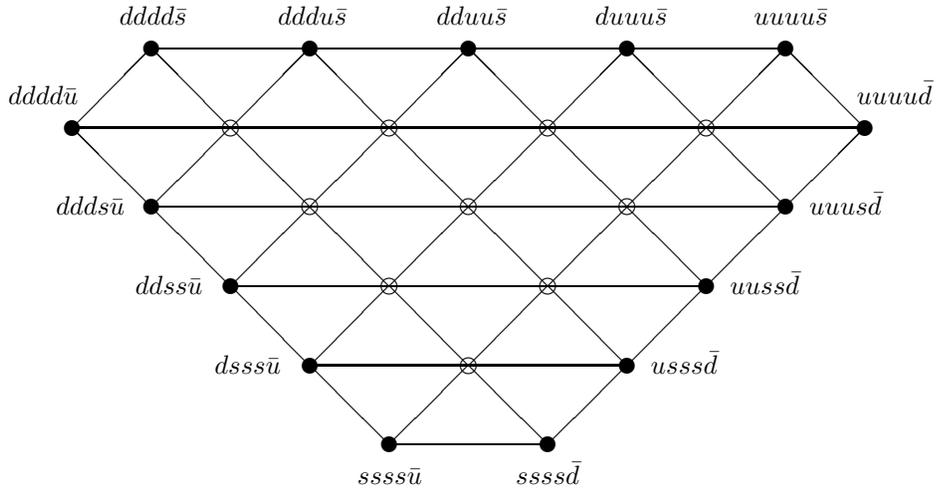

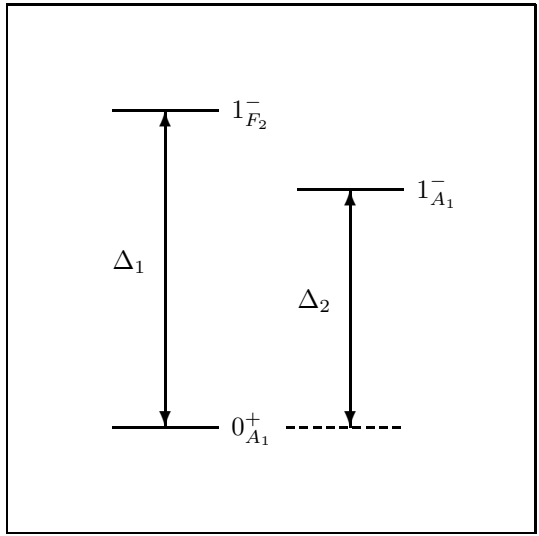
\begin{figure}
\centering
\setlength{\unitlength}{1.0pt}
\begin{picture}(200,200)(0,0)
\thinlines
\put(  0,  0) {\line(1,0){200}}
\put(  0,200) {\line(1,0){200}}
\put(  0,  0) {\line(0,1){200}}
\put(200,  0) {\line(0,1){200}}
\thicklines
\put( 40, 40) {\line(1,0){ 40}}
\put( 40,160) {\line(1,0){ 40}}
\put(110,130) {\line(1,0){ 40}}
\put( 60, 40) {\vector(0, 1){120}}
\put( 60,160) {\vector(0,-1){120}}
\put( 40,100) {$\Delta_1$}
\put(130, 40) {\vector(0, 1){90}}
\put(130,130) {\vector(0,-1){90}}
\multiput(106,40)(5,0){9}{\line(1,0){3}}
\put(110, 85) {$\Delta_2$}
\put( 85, 37) {$0^+_{A_1}$}
\put( 85,157) {$1^-_{F_2}$}
\put(155,127) {$1^-_{A_1}$}
\end{picture}
\vspace{15pt}
\caption[]{Orbital excitations of the pentaquark up to $N=1$ quantum. 
The states are labeled by angular momentum, parity and ${\cal T}_d$ 
symmetry $L^p_t$.} 
\label{orbex}
\end{figure}

\end{document}